\documentclass[pra,twocolumn]{revtex4-1}
\usepackage{dcolumn}% Align table columns on decimal point
\usepackage{bm}% bold math
\usepackage{amsmath,amssymb}
\usepackage{bm}
\usepackage{graphicx}
\usepackage{ascmac}
\usepackage[latin1]{inputenc}
\usepackage{braket}
\usepackage{txfonts}
\usepackage{mathrsfs}
\usepackage{color}
\usepackage[bottom]{footmisc}
\def\be{\begin{equation}}  % \begin{equation}
\def\ee{\end{equation}}     % \end{equation}
\def\I{\mathrm{i}}  %imaginary number 
\def\tr{\mathrm{tr}}
\def\e{\mathrm{e}} 
\def\Ls{{L}}
\def\Js{{J}}
\def\p{| \vec{p} \, |}
\def\Lslmbd{{L^{\Lambda}_{\: }}}
\def\Jslmbd{{J^{\Lambda}_{\, \mathrm{S}}}}
\def\Jslmbd{{J^{\Lambda}_{\, }}}
\def\Jsrel{{J^{\mathrm{rel}}}}
\def\erfc{\mathrm{erfc}} 
\begin{document}
\title{Effect of Wigner rotation on estimating unitary-shift parameter of relativistic spin-1/2 particle}
\author{Shin Funada}
\author{Jun Suzuki}
\affiliation{%
Graduate School of Informatics and Engineering, The University of Electro-Communications, 
1-5-1 Chofugaoka, Chofu-shi, Tokyo, 182-8585 Japan
}%

\date{\today}

\begin{abstract}
We obtain the accuracy limit for estimating the expectation value of the position of a relativistic particle for an observer moving along one direction at a constant velocity. We use a specific model of a relativistic spin-1/2 particle described by a gaussian wave function with a spin down in the rest frame. To derive the state vector of the particle for the moving observer, we use the Wigner rotation that entangles the spin and the momentum of the particle. Based on this wave function for the moving frame, we obtain the symmetric logarithmic derivative (SLD) Cram\'er-Rao bound that sets the estimation accuracy limit for an arbitrary moving observer. It is shown that estimation accuracy decreases monotonically in the velocity of the observer when the moving observer does not measure the spin degree of freedom. This implies that the estimation accuracy limit worsens with increasing the observer's velocity, but it is finite even in the relativistic limit. We derive the amount of this information loss by the exact calculation of the SLD Fisher information matrix in an arbitrary moving frame.
\end{abstract}
\maketitle 
%=====================================================================
\section{Introduction}
%=====================================================================　
Relativistic quantum information theory brings the new direction of research in physics. 
The significance of the effect of the relativity on the quantum state is that the state vector for a moving observer 
changes depending on the motion of the observer while the physical state in the rest frame remains the same. 
As a natural consequence, information which the moving observer obtains changes depending on the motion 
of the moving observer since the state vector changes.  
There are studies of quantum information theory with the relativity being taken into account. 
The studies in the realm of the relativistic quantum information have increased in number in past. 
Here, we briefly list some of them. 
Firstly, information paradox about black holes is now formulated in the framework of information theory, see recent reviews~\cite{harlow,maldacena}. 
Secondly, quantum information in non-inertial frame was investigated~\cite{alsing3,alsing2,bruschi,hosler,yao,alsing2}. 
Thirdly, the effect of the relativity on the the Bell's inequality. 
The degree of the violation of the Bell's inequality was investigated~\cite{ahn,terashima2,terashima,moon,moradi,caban}. 
The entropy changes due to the relativistic effect~\cite{peres2} and its effect on the Bell's inequality are 
also studied, which was initiated in~\cite{terno}. 

Among these early studies about the relativity and the quantum information, 
the papers~\cite{terashima,terashima2,ahn,alsing,peres3} brought the use of the Wigner rotation~\cite{halpern,weinberg} into the realm of quantum information. 
As other examples, 
the Wigner rotation is used to discuss the limitation given by a quantum entropy in the relativity domain
~\cite{peres2}.  
The entanglement~\cite{jordan,alsing,gingrich,lee,pachos,lamata,friis,castro} and Bell's inequality~\cite{terashima,ahn,kim} are also discussed by using the Wigner rotation. 
The essence of the Wigner rotation is that it `rotates' the spin of the relativistic particle 
by the angle, which is a function of the momentum of the particle. 
Thus, the spin and the momentum couple in a non-trivial way that the Wigner rotation gives. 

Based on the previous investigations in relativistic quantum information theory, 
it is natural to pose a question: what is the effect of the Wigner rotation for parameter estimation about  quantum states? 
To phrase it differently, we ask: how does estimation accuracy change for a moving observer? 
However, to the best of our knowledge, there has not existed a study about 
the change in estimation accuracy that a moving observer undergoes. 
We demonstrate how estimation accuracy changes for the moving observer 
in the framework of the quantum estimation theory~\cite{holevo,helstrom}. 
To obtain the limit of estimation accuracy as a function of the moving observer's velocity, 
we utilize the quantum Fisher information matrix 
which enables us to quantify the accuracy limit. 
Among those quantum Fisher information matrices, 
we consider the symmetric logarithmic derivative (SLD) Fisher information matrix 
as an indicator of estimation accuracy. 
As the main result, we obtain the analytical expression of the SLD Fisher information matrix 
for an arbitrary moving observer as an integral form Eq.~\eqref{eq:Jinverse}. 
This then sets the estimation accuracy limits 
between the observers in the rest frame and in the moving frame. 
To illustrate our result, we plot the relativistic effect on estimation accuracy in Fig.~\ref{fig4}. 
Estimation accuracy obtained by the SLD Fisher information matrix is 
finite even at the relativistic limit where the velocity $V$ approaches the speed of light. 
This suggests that estimation accuracy remains finite at the relativistic limit. 

As for the model, we set up a specific pure-state model that describes a single spin-1/2 particle. 
A parametric model is defined by a two-parameter unitary shift model. 
We next consider an observer moving at a constant velocity in one direction with respect to the rest frame.  
The moving observer then makes a measurement to estimate the parameters encoded in the state without accessing the spin degree of freedom. 
Thus, our parameter model in the moving frame is given by the Wigner rotation followed by 
the partial trace over the spin. 
In our study, the parameters correspond to the expectation value for the position of the particle. 
We investigate how estimation accuracy for the moving observer changes as a function of the velocity. 
We evaluate the limits for the mean square error (MSE) upon 
estimating the expectation value of the position operator by the SLD Cram\'er-Rao (CR) bound. 
We obtain analytically how much the accuracy decreases as a function of the velocity of the moving observer. 

Before closing the introduction, let us briefly remark on the earlier works of estimation theory 
in the relativistic domain. A seminal paper~\cite{braunstein} triggers the study of the parameter 
estimation in the relativity domain. The result gives an insight into the quantum estimation theory in the relativistic domain.  
In their work, the authors derive an uncertainty relation based on 
the restriction provided by the Lorentz invariance.  
They do not consider the change in estimation accuracy or the uncertainty relation by the Lorentz transformation.  
The other instance of parameter estimation about a relativistic quantum state is 
known as relativistic quantum metrology~\cite{ahmadi2,ahmadi,tian,liu}. 
However, these studies do not address the question proposed in this paper. 

The outline of this paper is as follows. 
In Sec.~\ref{sec:Model}, our model is explained. The state in the rest frame is given.  The state in the moving frame is 
derived by applying the Wigner rotation to the state in the rest frame. 
In Sec.~\ref{sec:Estimation}, we explain our parameter estimation which is given by 
the wave function derived by the Wigner rotation. 
We evaluate first the SLD CR bound in such a case that 
the moving observer does not have information about the degree of freedom of spin state. 
We also make comments about other two possible cases. 
The SLD CR bounds are investigated by multi-parameter estimation and given in analytical forms. 
In Sec.~\ref{sec:Discussion}, we discuss how the Wigner rotation changes the wave function and 
how it gives rise to the loss of information. 
Sec.~\ref{sec:Conclusion} gives the conclusion. 
Appendix~\ref{sec:Wigner} summarizes well known facts about the Wigner rotation for a massive 
spin-1/2 particle. Most of the technical calculations are presented in the remaining Appendices. 
%
%=====================================================================
\section{Model} \label{sec:Model}
%=====================================================================
We assume that an observer moves along the $z$ axis with a constant velocity $V$. 
We choose the $z$ direction as the moving direction, 
because we expect that this direction gives the most significant change in 
the rotation of spin as a massive relativistic spin-$1/2$ particle on the $x$-$y$ plane~\cite{terashima}. 
We use the natural unit, i.e., $\hbar=1$ and $c=1$ unless otherwise stated. 
The mass of the particle is $m$. As a metric tensor $g_{\mu \nu}$, we choose 
$g_{\mu \nu}= (+1, -1, -1, -1)$.
\subsection{State in the rest frame}
The wave function of the particle is set as a gaussian function of $x$ and $y$ with a plane wave in 
the $z$ coordinate. For simplicity, we set the wave number, or the momentum along the $z$ direction as zero. 
To apply the Wigner rotation as described in~\cite{halpern, weinberg}, 
we mainly use the momentum representation in the following discussion. 

The state of the particle is in a known pure state called a reference state. 
The reference state $\rho_0$ in the rest frame is 
\begin{align}
\rho_0&=\ket{\Psi_\downarrow} \bra{\Psi_\downarrow}, \label{eq:rho0}\\
\ket{\Psi_\downarrow}&= \int d^3p \, \varphi_0(p^1) \varphi_0(p^2)\delta(p^3) \ket{\vec{p}, \downarrow},  \nonumber
\end{align}
where $\delta(p^{3})$ denotes the Dirac delta function to represent the plane wave in the $z$ direction. 
The momentum vector $\vec{p}$ is a spatial part of the four-momentum $p^\mu$, i.e., $\vec{p}=(p^1,p^2,p^3)$. 
The state vectors $ \ket{\vec{p}, \downarrow}$ and $ \ket{\vec{p}, \uparrow}$ are the momentum eigenstates with down and up spins, respectively. 
The $\varphi_0(p)$ is defined by the gaussian function as
\be
\varphi_0(p)=
\frac{\kappa^{1/2}}{\pi^{1/4}}\, \e^{- \frac12\kappa^2 p^2}. \label{eq:psi_0}
\ee
The $\kappa$ determines the spread of the wave function in the coordinate representation, i.e., the spread of the wave function in the coordinate representation becomes broader as $\kappa$ increases. 
The spread $\kappa$ is a quantity that an experimenter chooses at his/her will. 

A quantum parametric model is defined by a two-parameter unitary model as
\be\label{eq:model_rest}
\mathcal{M}_{\rm rest}= \big\{ \rho_\theta \,\big|\, \theta = (\theta_1, \theta_2) \subset \mathbb{R}^2 \big\}, 
\ee
where $\rho_\theta$ is generated by the momentum operators in the $x$ and $y$ direction, $\hat{p}^1$ and $\hat{p}^2$, respectively, 
\be
\rho_\theta=U(\theta) \rho_0 U^\dagger (\theta)
=U(\theta) \ket{\Psi_\downarrow} \bra{\Psi_\downarrow}U^\dagger (\theta), \label{eq:rho_theta}
\ee
with
\be
U(\theta)=\e^{-\I \hat{p}^1 \theta_1 - \I \hat{p}^2 \theta_2}. \label{eq:unitary}
\ee
The operator $\hat{p}^i$ ($i=1,2$) are the momentum operator of $i$th component, i.e., 
$\hat{p}^i \ket{\vec{p}, \sigma} = p^i \ket{\vec{p}, \sigma}$, $(\sigma= \downarrow, \uparrow)$. 
Let us define a state vector $\ket{\Psi_\downarrow (\theta)}$ by 
\begin{align}
\ket{\Psi_\downarrow (\theta)} 
&=U(\theta) \ket{\Psi_\downarrow}\nonumber\\
&=\int d^3p \, \varphi_0(p^1)\varphi_0(p^2) \delta(p^3) \e^{-\I p^1 \theta_1 - \I p^2 \theta_2} \ket{\vec{p}, \downarrow}. \label{eq:Psi_theta}
\end{align}
Then, Eq.~\eqref{eq:rho_theta} is expressed as 
\be
\rho_\theta=\ket{\Psi_\downarrow (\theta)} \bra{\Psi_\downarrow (\theta)}.
\ee
The physical implication of the parameter $\theta$ is that it is the peak position of the wave function 
in the coordinate representation. 
Alternatively, we consider position operators $\hat{x}^j$, which are canonical conjugate of 
the momentum operators $\hat{p}^j$, ($j=1,2$)~\cite{comment}. 
From Eq.~\eqref{eq:unitary}, we have 
\be \label{eq:shiftX}
U(\theta) \hat{x}^j U^\dagger(\theta)= \hat{x}^j + \theta_j, \quad (j=1, 2). 
\ee
The unitary transformation $U(\theta)$ gives a shift by $\theta_j$ to a position operator $\hat{x}^j$. 
By assumption, we know the reference state $\rho_0$.  However, we do not know $\theta_1$ or $\theta_2$. 
We estimate the parameters $\theta_1$ and $\theta_2$ encoded in $\rho_\theta
=U(\theta) \ket{\Psi_\downarrow} \bra{\Psi_\downarrow} U^\dagger (\theta)$. 
By doing so, we have an estimate for the expectation value of the position operators $\hat{x}^1$ and $\hat{x}^2$ as seen in Eq.~\eqref{eq:shiftX}.  

The parametric model \eqref{eq:model_rest} in the rest frame is a classical model in the following sense. 
Firstly, two parameters are totally uncorrelated since the state vector \eqref{eq:Psi_theta} is also expressed as the tensor product form, 
\be
\ket{\Psi_\downarrow (\theta)}=\ket{\psi_{1} (\theta_{1})}\ket{\psi_{2} (\theta_{2})}\ket{p^{3}=0}\ket{\downarrow}, \nonumber
\ee
with 
\be
\ket{\psi_{j} (\theta_{j})}= \int dp^j \, \varphi_0(p^j)  \e^{-\I p^j \theta_j} \ket{p^j},\quad (j=1,2). \nonumber
\ee
Secondly, an optimal measurement to estimate $\theta_{j}$ is the position operator $\hat{x}^{j}$. 
Optimal measurements for $\theta_{1}$ and $\theta_{2}$ commute and hence we can simultaneously 
perform the optimal measurement. 
Thirdly, upon measuring the position operators, the measurement outcomes 
obey the independent classical gaussian distributions with the mean $(\theta_{1},\theta_{2})$ 
and their variances $(\kappa^{2}/2,\kappa^{2}/2)$. Thus, the optimal unbiased estimator is given by the sample mean. 

\subsection{Quantum Fisher information in the rest frame}
The symmetric logarithmic derivative (SLD) $\Ls_{j} (\theta)$ 
of the pure state model \eqref{eq:model_rest} is calculated, for example, 
by the method given in~\cite{fujiwara} as 
\be
\Ls_{j}(\theta)= 2 \partial_j [ \ket{\Psi_\downarrow(\theta)} \bra{\Psi_\downarrow(\theta)} ], \label{eq:rest_SLD}
\ee
where $\partial_j = \partial/\partial \theta_j$. 
By a direct calculation, we obtain the commutator of the SLDs as
\be
[\Ls_{1}(\theta), \, \Ls_{2}(\theta)]= 4 (\ket{\partial_1 \Psi_\downarrow(\theta)} \bra{\partial_2 \Psi_\downarrow(\theta)}
-\ket{\partial_2 \Psi_\downarrow(\theta)} \bra{\partial_1 \Psi_\downarrow(\theta)} ), \nonumber
\ee
where $\ket{\partial_j \Psi_\downarrow(\theta)}=\partial_j\ket{ \Psi_\downarrow(\theta)}$, ($j=1,2$). 
Similar notations will be used throughout the paper. 
We remark that the SLDs do not commute in this particular choice of SLDs. 

At first sight, this non-commutativity seems to contradict the fact that the parametric model in the rest frame is 
a classical one. 
A resolution is that the choice of the SLDs above is not unique \cite{fujiwara}. 
As an example, we have another choice of the SLDs, 
$\tilde{L}_{j}(\theta)$ ($j=1,2$) as follows. 
\begin{align}
\tilde{L}_{1}(\theta)&=2 \partial_1 ( \ket{\psi_{1} (\theta)} \bra{\psi_{1} (\theta)}) 
\otimes \mathrm{I}_2 \otimes \mathrm{I}_3 \otimes \ket{\downarrow} \bra{\downarrow} \label{eq:alt_L1}, \\
\tilde{L}_{2}(\theta)&=\mathrm{I}_1 \otimes 2 \partial_2 ( \ket{\psi_{2} (\theta)} \bra{\psi_{2} (\theta)})
\otimes \mathrm{I}_3 \otimes \ket{\downarrow} \bra{\downarrow}, \label{eq:alt_L2} 
\end{align}
where
\begin{align}
I_k&= \int dp^k \, \ket{p^k}  \bra{p^k}, \quad (k=1, \, 2, \, 3). \nonumber 
\end{align}
These SLDs $\tilde{L}_{j}(\theta)$ satisfy the definition of SLD and indeed they do commute each other.  

The SLD Fisher information matrix $\Js(\theta)=[ \Js_{jk}(\theta)] $ is obtained by the formula in~\cite{fujiwara} as
\begin{align}
\Js_{jk} 
&=4(\braket{\partial_j  \Psi_\downarrow(\theta) | \partial_k \Psi_\downarrow(\theta)} \nonumber \\
&+\braket{ \Psi_\downarrow(\theta) | \partial_j \Psi_\downarrow(\theta)}
\braket{ \Psi_\downarrow(\theta) | \partial_k \Psi_\downarrow(\theta)}). \nonumber 
\end{align}
In the following discussion, we drop $\theta$ in the SLD Fisher information matrix, 
because $\Js$ is independent of $\theta$ due to the unitarity of the model. 
By a straightforward calculation involving the standard gaussian integrals, we have
\be
 \Js_{jk}= \frac{2}{\kappa^2} \delta_{jk}, \quad (j,k=1,2). \label{eq:rest_SLD}
\ee 
The alternative SLDs $\tilde{L}_{j}(\theta)$ Eqs.~\eqref{eq:alt_L1} and~\eqref{eq:alt_L2} give  
the same SLD Fisher information matrix Eq.~\eqref{eq:rest_SLD}.
The inverse of the SLD Fisher information matrix $\Js^{-1}=[ \Js_{j k}^{\:-1}]$ is also diagonal as follows.
\be
\Js_{jk}^{\; -1}= \frac{\kappa^2}{2}\delta_{jk}. 
\label{eq:J_rest}
\ee
The SLD CR inequality is expressed as
\be
V \geq J^{-1}, \nonumber
\ee
where $V=[V_{jk}]$ is the mean square error (MSE) matrix. 
With Eq.~\eqref{eq:J_rest}, we have 
\be
V_{11} \geq \frac{\kappa^2}{2}, \quad V_{22} \geq \frac{\kappa^2}{2}. \label{SLD_CRinequality_noLB}
\ee
The estimation accuracy limit regarding the expectation value of the position operator 
is proportional to $\kappa^2$ which determines the spread of the wave function in 
the coordinate representation. 
It is easy to see $\Js^{\; -1}$ approaches the zero matrix as $\kappa \rightarrow 0$. 
At the limit of $\kappa \rightarrow 0$, the wave function in the coordinate representation 
becomes the Dirac delta function. This allows us to estimate the parameter $\theta$ without any error. 
\subsection{State in a moving frame}
We next consider an observer moving along the $z$ axis with respect to the rest frame. 
A Lorentz transformation $\Lambda$ from the rest frame to this moving frame is 
\begin{align}
\Lambda&=
\left(
\begin{array}{cccc}
 \cosh \chi  & 0 & 0 & -\sinh \chi  \\
 0 & 1 & 0 & 0 \\
 0 & 0 & 1 & 0 \\
 -\sinh \chi  & 0 & 0 & \cosh \chi  \\
\end{array}
\right),  \\
\cosh \chi &= \frac{1}{\sqrt{1-V^2}}, \quad \sinh \chi = \frac{V}{\sqrt{1-V^2}}. \label{eq:Lambda}
\end{align}
$V$ is a velocity of the observer moving along the $z$ axis. 
By this Lorentz transformation, the momentum of the particle is transformed as in classical physics. 
The spatial part of the four-momentum, $\Lambda \vec{p}$ is given by 
\begin{align}
\Lambda \vec{p} 
&= \left(\sum_{\mu=0}^3 \Lambda^1_{\: \mu} p^\mu,\, \sum_{\mu=0}^3 
\Lambda^2_{\: \mu} p^\mu,\, \sum_{\mu=0}^3 \Lambda^3_{\: \mu} p^\mu\right) \nonumber \\
&=(p^1, p^2, -p^0 \sinh \chi), 
\end{align}
where $p^0=\sqrt{m^2+ \p^2}$. See for example~\cite{halpern}.

For a relativistic spin-1/2 particle, the Lorentz transformation $\Lambda$ also gives rise to a unitary transformation $U(\Lambda)$ acting on the state vector. 
This is described by the Wigner rotation~\cite{halpern,weinberg} (See a short summary in Appendix~\ref{sec:Wigner}.). 
In our model, the state vector in the rest frame is in a spin down state, $\ket{\Psi_{\downarrow}(\theta)}$. 
The state vector $\ket{\Psi_{\downarrow}(\theta)}$ is transformed to $\ket{\Psi^\Lambda (\theta)}$ as 
\be \label{eq:Psilambda}
\ket{\Psi^\Lambda (\theta)} =U(\Lambda) \ket{\Psi_\downarrow (\theta)}
= \sum_{\sigma=\downarrow, \uparrow} \ket{\Psi^\Lambda_{\: \sigma}(\theta)}. 
\ee
We remark here that $\ket{\Psi^\Lambda_{\: \sigma}(\theta)}, \: (\sigma= \downarrow, \uparrow)$ are not normalized. 
It is convenient to express the state vectors 
$\ket{\Psi^\Lambda_{\: \sigma}(\theta)}$, $(\sigma= \downarrow, \uparrow)$ as
\begin{align}
\ket{{\Psi^\Lambda_{\: \sigma}(\theta)}}= \ket{\psi^\Lambda_{\: \sigma}(\theta)} \ket{\sigma}. \nonumber 
 \end{align}
The explicit form of $\ket{{\Psi^\Lambda_{\: \sigma}(\theta)}}$ is given by 
\begin{align}
 \ket{\psi^\Lambda_{\: \sigma}(\theta)}&= \int d^3p \sqrt{\frac{(\Lambda p)^0}{p^0}} F_{\theta, \, \sigma}({p}^1, {p}^2) \delta(p^3) \ket{\Lambda \vec{p}}, 
  \label{eq:Psi_pi}\\
F_{\theta, \, \downarrow}(p^1, p^2)
&=\varphi_0(p^1)\varphi_0(p^2) \e^{- \I p^1 \theta_1- \I p^2 \theta_2} \cos \frac{\alpha(\p)}{2}, \label{eq:F1} \\
F_{\theta, \, \uparrow}(p^1, p^2)
&=-\varphi_0(p^1)\varphi_0(p^2)  \e^{- \I p^1 \theta_1- \I p^2 \theta_2} \e^{\I \phi(p^1, \, p^2)}\sin \frac{\alpha(\p)}{2}, \label{eq:F2} \\
\p &=\sqrt{(p^1)^2+(p^2)^2}, \nonumber \\
\e^{\I \phi(p^1, \, p^2)}&= \frac{p^1}{\p}+ \I \frac{p^2}{\p}, \nonumber \\
\cos \alpha(\p)&=\frac{\sqrt{m^2+ \p^2} + m \cosh \chi}{\sqrt{m^2+ \p^2} \cosh \chi + m }, \label{eq:cosbeta} \\
\sin \alpha(\p)&= - \frac{\p \sinh{\chi}}{\sqrt{m^2+ \p^2} \cosh \chi + m }.  \label{eq:sinbeta}
\end{align}
In the expressions above, $m$ denotes the mass of the spin-1/2 particle. 

The Lorentz boost gives a non-zero probability density of spin up state as shown in Eq.~\eqref{eq:F2}.  
This makes the particle spin `rotate', and hence is called the Wigner rotation. 
Detailed derivations of Eqs.~\eqref{eq:Psilambda},~\eqref{eq:Psi_pi},~\eqref{eq:F1}, and~\eqref{eq:F2} are given 
in~Appendix~\ref{sec:Wigner}.

We remark that the states $\ket{\Psi^\Lambda(\theta)}$ 
are entangled with respect to the momentum and the spin degrees of freedoms. 
For the observer moving along the $z$ axis, the spin has a component of spin up which is none at the rest frame, 
i.e., the spin rotates as the observer moves. 
%
%=====================================================================
\section{Parameter estimation: moving frame} \label{sec:Estimation} 
%=====================================================================
We are now in position to discuss parameter estimation in the moving frame. 
Suppose that a moving observer wishes to estimate the parameter $\theta$ encoded in the state Eq.~\eqref{eq:Psilambda}. 
The system under discussion has two different degrees of freedoms. 
One is continuous one describing the wave function, and the other is the spin. 
It is natural to measure the continuous degree of freedom to estimate the parameter as 
the observer does not know whether s/he is in a moving frame or not. 
In this setting, the moving observer does not have access to the spin degree of freedom. 
Then, our parametric model is given by tracing out the spin of from the pure state Eq.~\eqref{eq:Psilambda}. 

As comparison, we also give a short account on other possible cases. 
The first is when the moving observer measures the both degrees of freedoms. 
This will be discussed in Sec.~\ref{sec:QFI_boost}. 
The other case is when the spin of the particle is measured only, which will be given in Sec.~\ref{sec:model_spin}. 
%=====================================================================
\subsection{Invariance of quantum Fisher information after the Lorentz boost}\label{sec:QFI_boost}
%=====================================================================
We first consider the situation where the moving observer measures the whole state Eq.~\eqref{eq:Psilambda}. 
The parametric model for this case is defined as follows.
\be \label{def:model_boost}
\mathcal{M}_{\rm boost}= \big\{\ket{{\Psi^\Lambda (\theta)}}\bra{{\Psi^\Lambda (\theta)}} \,\big| \, \theta=(\theta_1, \theta_2) \in \mathbb{R}^2 \big\}.
\ee
It is clear that this model is unitary equivalent to the model in the rest frame, since the difference is 
only given by the unitary transformation $U(\Lambda)$. 
To phrase it differently, we can regard the model after the Lorentz boost in the different representation. 
Therefore, the SLD Fisher information matrix is exactly same as in the rest frame, Eq.~\eqref{eq:J_rest}. 
While this is true mathematically, the physical meanings of these two models are different. 

Let us further elaborate on physics of the two models; the one in the rest frame Eq.~\eqref{eq:model_rest} and the other Eq.~\eqref{def:model_boost} in the moving frame. 
The unitary transformation $U(\Lambda)$ which defines the Wigner rotation is 
parameter independent, and hence, two parametric models are equivalent. 
Yet, the significance of the Lorentz transformation is that $U(\Lambda)$ depends on the velocity $V$ of the moving observer with respect to the rest frame. 
The resulting state-vector after the Lorentz boost Eq.~\eqref{eq:Psilambda} indeed depends on $V$ in a non-trivial manner. Furthermore, the wave function in the moving frame is no longer described by the simple gaussian wave function as given in Eqs.~\eqref{eq:F1} 
and~\eqref{eq:F2}. 
In particular, the two parameters $\theta_{1}$ and $\theta_{2}$ are not described by 
a tensor product of two independent parametric models as in the rest frame. 
Nevertheless, we can formally express an optimal measurement for the model after the Lorentz boost by the pair of observables 
\begin{equation*}
U(\Lambda)\hat{x}^{j}U^{\dagger}(\Lambda),\quad (j=1,2), 
\end{equation*}
which obviously commute each other. 
We will not give further analysis on these observables, but it is obvious that 
experimental implementation of this optimal measurement is much more complex. 
It may not be feasible as it will depend on the velocity $V$. 
%=====================================================================
\subsection{Parametric model in the moving frame} \label{sec:moving_frame}
%=====================================================================
We now analyze the parametric model when the moving observer does not measure the spin of the particle. 
By taking the partial trace over the spin $\sigma$, we have 
\begin{align}
 \rho^\Lambda(\theta)
 &= \tr_{\mathrm{\sigma}} \ket{{\Psi^\Lambda(\theta)}} \bra{ {\Psi^\Lambda(\theta)}} \nonumber \\
 &= \sum_{\sigma=\downarrow, \uparrow} 
 \braket{\sigma | \Psi^\Lambda(\theta)} \braket{\Psi^\Lambda(\theta) | \sigma} \nonumber \\
 &=\sum_{\sigma=\downarrow, \uparrow} \ket{\psi^\Lambda_{\: \sigma}(\theta)}\bra{\psi^\Lambda_{\: \sigma}(\theta)}.  
\nonumber 
\end{align}
With this $ \rho^\Lambda(\theta)$, we define the parametric model of interest as 
\be \label{def:model_moving}
\mathcal{M}^{\Lambda}= \big\{  \rho^\Lambda(\theta) \, \big| \, \theta=(\theta_1, \theta_2) \in \mathbb{R}^2 \big\}.
\ee

As noted before, the state vectors $\ket{\psi^\Lambda_{\: \sigma}(\theta)}$ are unnormalized. 
By using the normalized state vector $\ket{\bar{\psi}^\Lambda_{\: \sigma}(\theta)}$ defined by
\be
\ket{\bar{\psi}^\Lambda_{\: \sigma}(\theta)}=\frac{\ket{\psi^\Lambda_{\: \sigma}(\theta)}} 
{\sqrt{\braket{\psi^\Lambda_{\: \sigma}(\theta) | \psi^\Lambda_{\: \sigma}(\theta)}}}, \nonumber
\ee
we  write $\rho^\Lambda(\theta)$ as a convex combination of 
two pure states $ \ket{\bar{\psi}^\Lambda_{\: \downarrow}(\theta)}\bra{\bar{\psi}^\Lambda_{\: \downarrow}(\theta)}$ and 
$ \ket{\bar{\psi}^\Lambda_{\uparrow}(\theta)}\bra{\bar{\psi}^\Lambda_{\uparrow}(\theta)}$, i.e., 
\be
\rho^\Lambda(\theta)=
\frac{1 }{2} (1 + \xi) \ket{\bar{\psi}^\Lambda_{\: \downarrow}(\theta)}\bra{\bar{\psi}^\Lambda_{\: \downarrow}(\theta)}
+\frac{1 }{2} (1 - \xi) \ket{\bar{\psi}^\Lambda_{\: \uparrow}(\theta)}\bra{\bar{\psi}^\Lambda_{\: \uparrow}(\theta)}. 
\nonumber
\ee

Let us evaluate the inner products $\braket{\psi^\Lambda_{\: \sigma}(\theta) | \psi^\Lambda_{\: \sigma}(\theta)}$ to analyze the amplitudes of the each spin state. 
From Eqs.~\eqref{eq:Psi_pi},~\eqref{eq:F1}, and~\eqref{eq:F2}, the inner products are written as
\begin{align}
\braket{\psi^\Lambda_{\: \downarrow}(\theta) | \psi^\Lambda_{\: \downarrow}(\theta)}=\frac{1}{2}( 1 + \xi), \label{eq:rho_1} \\
\braket{\psi^\Lambda_{\: \uparrow}(\theta) | \psi^\Lambda_{\: \uparrow}(\theta)}=\frac{1}{2}( 1 - \xi), \label{eq:rho_2} 
\end{align}
where
\be
\xi = \kappa^2 \int_0^\infty dp \, (2p)\,
\frac{\sqrt{m^2+p^2} \sqrt{1-V^2}+m}{\sqrt{m^2+p^2}+ m\sqrt{1-V^2}} \,\e^{-{\kappa}^2 p^2} . 
\label{eq:xi2} 
\ee

The $\xi$ is an indicator of the spin rotation by the Lorentz boost as seen in Eqs.~\eqref{eq:rho_1} and~\eqref{eq:rho_2}.  
As the result, it depends only on the observer's velocity $V$. 
The smaller $\xi$ becomes, the larger the amplitude of the spin up state. Therefore, the spin rotates. 

\begin{figure}[t]
\begin{center}
\includegraphics[width=7cm]{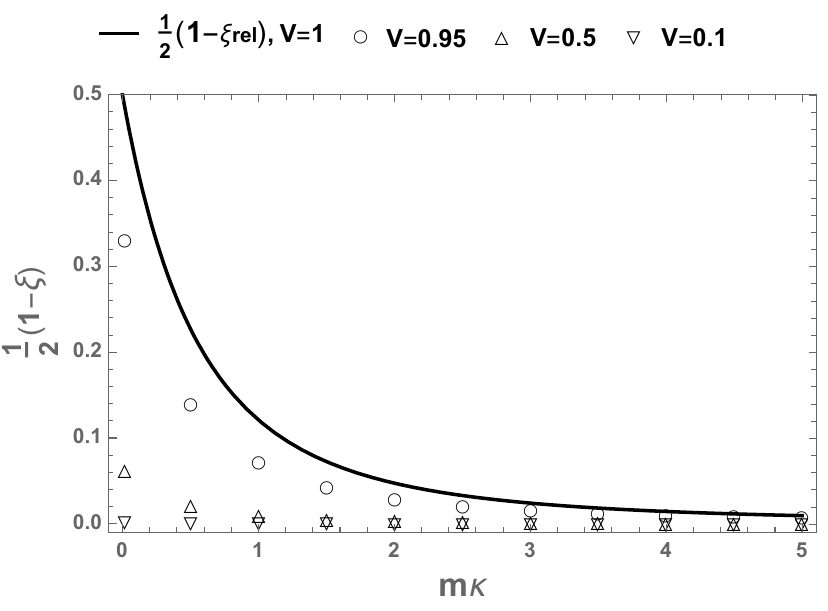}
\caption{Numerically calculated 
$(1-\xi)/2$ as a function of $m \kappa$ at  $V=1, 0.95, 0.5,$ and $0.1$. 
The set of the velocity $V$ is chosen differently to make the distance between the plots more even.}
\label{fig1}
\end{center}
\end{figure}

At any given $\kappa$, the $\xi$ takes its maximum value $1$ at $V=0$ which means no spin rotation. 
It takes its minimum value $\xi_\mathrm{rel}$ 
in the relativistic limit of $V \rightarrow1$ which corresponds to $V \rightarrow c$ in the standard unit.  
An explicit expression of $\xi_\mathrm{rel}$ is 
\be
\xi_\mathrm{rel}=\sqrt{\pi } m  \kappa \,  \e^{ m^2\kappa ^2} \erfc(m \kappa ), \label{eq:xi_min}
\ee 
where $\erfc(x)$ is the complementary error function defined by
\be
\erfc(x) 
= \frac{2}{\sqrt{\pi}} \int_x^\infty dt\,\e^{-t^2}. \nonumber 
\ee
The derivations of Eqs.~\eqref{eq:rho_1},~\eqref{eq:rho_2},~\eqref{eq:xi2}, and~\eqref{eq:xi_min} 
are given in~Appendix~\ref{sec:xi}.
The probability for the spin up state reaches its maximum $1/2$ at the limit of $\kappa \rightarrow 0$ and 
at the relativistic limit. 
Figure~\ref{fig1} shows the probability of the spin up state \\
$\braket{\psi^\Lambda_{\: \uparrow}(\theta) | \psi^\Lambda_{\: \uparrow}(\theta)} = (1-\xi)/2$
as a function of $m \kappa$ at $V=0.95, 0.5,$ and $0.1$.  
The set of the velocities V is chosen differently to make the distance between the plots more even. 
Figure~\ref{fig1} shows its maximum $(1- \xi_\mathrm{rel})/2$ as well. 

Let us analyze these state vectors in the coordinate representation. 
We define the wave function of a particle with spin up in the coordinate representation $\psi^\Lambda_{\: \uparrow}(x)$ by 
\be
\psi^\Lambda_{\: \uparrow}(x)= \braket{x | \bar{\psi}^\Lambda_{\: \uparrow}(\theta)} \big|_{\theta=0}. \nonumber
\ee
A derivation of its explicit expression is given in Appendix~\ref{sec:wavefunction}.
Figure~\ref{fig2} shows numerically calculated densities $| \psi^\Lambda_{\: \uparrow}(x) |^2$ for $\kappa = 0.1$ as a function of the position $x^1$ for $V=0.99,~0.98,~0.7,$ and 0.1.
For simplicity, we set $(\theta_1, \theta_2)=(0,0)$ and $x^2, x^3=0$.  
It is worth noting that the peak of the spin up wave function $\psi^\Lambda_{\: \uparrow}(x)$ is no longer at $x^1=\theta^1=0$. 
To see the dependence of the observer's velocity $V$ on the peak position, we numerically calculate the derivative of 
$| \psi^\Lambda_{\: \uparrow}(x) |^2$.  Figure~\ref{fig3} shows the derivative of $| \psi^\Lambda_{\: \uparrow}(x) |^2$ as 
a function of position. In this figure, we set $(\theta_1, \theta_2)=(0,0)$ as well for simplicity. We observe that the faster the observer moves, the further the peak position moves away from $x^1=\theta^1$.
These numerically verified facts indicate that the parametric model Eq.~\eqref{def:model_moving} is   
a convex mixture of two pure state models. One is centered at $\theta$, and the other is centered at $\theta$+some amount. If one performs the position measurement, the resulting probability distribution is thus given by a convex mixture of two distributions with different locations of the peak. 
This finding naturally invites us to say that estimation accuracy gets worse for the moving observer. 
\begin{figure}[t]
\begin{center}
\includegraphics[width=7cm]{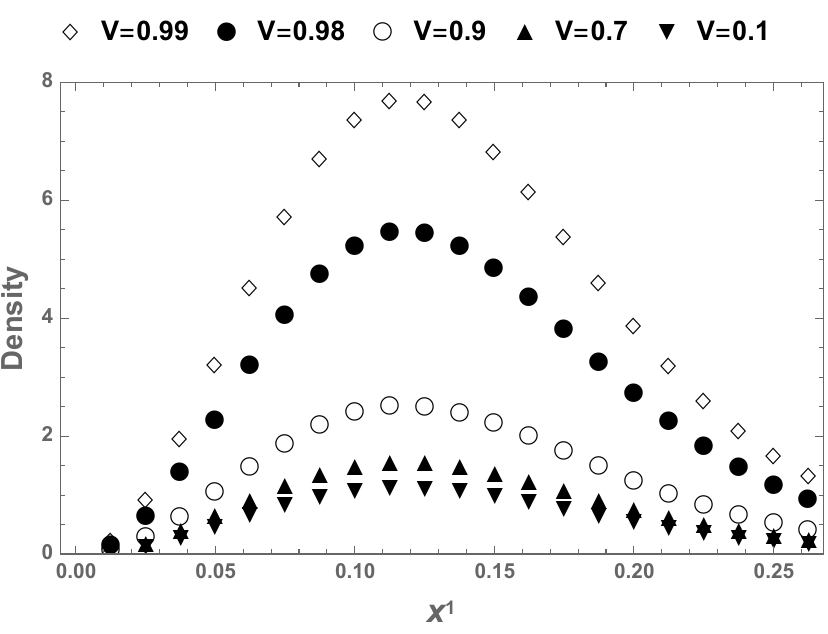}
\caption{Numerically calculated probability density $| \psi^\Lambda_{\: \uparrow}(x) |^2$ for $\kappa=0.1$ as a function of $x^1$ at  
$V=0.99, 0.98, 0.9, 0.7$ and $0.1$.}
\label{fig2}
\end{center}
\end{figure}
\begin{figure}[t]
\begin{center}
\includegraphics[width=7cm]{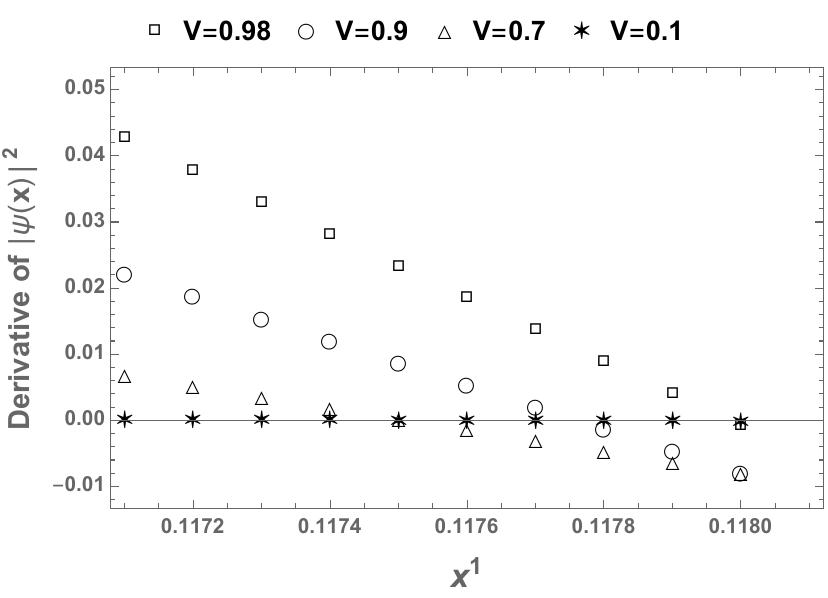}
\caption{Numerically calculated derivative of the probability density $| \psi^\Lambda_{\: \uparrow}(x) |^2$ as a function of $x^1$ at  $V=0.98, 0.9, 0.7$ and $0.1$.}
\label{fig3}
\end{center}
\end{figure}
\subsection{Quantum Fisher information matrix in the moving frame}
The SLD Fisher information matrix $\Jslmbd= [\Jslmbd_{jk}]$ for the model \eqref{def:model_moving} is calculated as follows.
\be
\Jslmbd_{jk}=\frac{2}{\kappa^2} (1- 2 \kappa^{2} \eta^2)
\delta_{jk}. \label{eq:Jinverse}
\ee
where
\be
\eta=- \int_{-\infty}^\infty \int_{-\infty}^\infty dp^1 dp^2 \frac{(p^1)^2}{\p} [\varphi_0 (p^1) \varphi_0(p^2)]^2 \sin \alpha(\p). \label{eq:eta}
\ee
A detailed explanation is given in Appendix~\ref{sec:SLD}. 
From Eq.~\eqref{eq:Jinverse}, we have the SLD CR inequality as follows.
\be
V_{11} \geq \frac{\kappa^2}{2} \frac{1}{1- 2 \kappa^{2} \eta^2}, \quad 
V_{22} \geq \frac{\kappa^2}{2} \frac{1}{1- 2 \kappa^{2} \eta^2}. \label{SLD_CRinequality_LB}
\ee
As given in Appendix~\ref{sec:J_ratio}, the denominator in Eq.~\eqref{SLD_CRinequality_LB}, $1- 2 \kappa^{2} \eta^2 $ is positive, 
and hence the accuracy limits for $V_{11}$ and $V_{22}$ are always finite. 

\begin{figure}[t]
\begin{center}
\includegraphics[width=7cm]{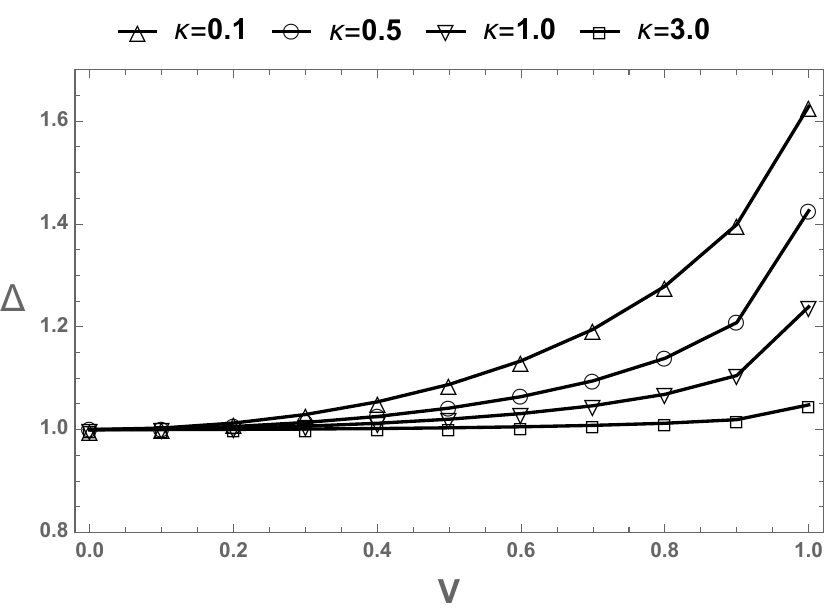}
\caption{Numerically calculated ratio $\Delta(V)$ as a function of $V$, the velocity of the moving observer at  $\kappa=0.1, 0.5, 1.0$ and $3.0$.}
\label{fig4}
\end{center}
\end{figure}

By comparing the SLD CR inequalities for the rest frame Eq.~\eqref{SLD_CRinequality_noLB} and 
Eq.~\eqref{SLD_CRinequality_LB}, we see how much estimation accuracy is affected 
by the Lorentz boost.  As an indicator, we take up the ratio of the $(1,1)$ components 
of $(\Jslmbd)^{-1}$ and $(\Js)^{-1}$. 
We define the ratio $\Delta(V)$ by
\be
\Delta(V) =\frac{[ (\Jslmbd)^{-1}]_{11}}{[(\Js)^{-1}]_{11}}=\frac{1}{1- 2 \kappa^{2} \eta^2}. \nonumber 
\ee
By definition, $\Delta(0)=1$ for the rest frame. 
The ratio $\Delta(V)$ quantifies the amount of information loss due to the Lorentz boost. 
If it is larger, the moving observer can only estimate the parameter less accurately when compared to the rest frame. 
Figure~\ref{fig4} shows the ratio $\Delta(V)$ as a function  
the moving observer's velocity $V$ at the different spreads of the wave function $\kappa$ = 0.1, 0.5, 1.0, and 3.0. 
The set of the spreads $\kappa$ is chosen to make the distance between the plots more even. 

From Eqs.~\eqref{eq:sinbeta} and~\eqref{eq:eta}, 
$\kappa \eta $ is expressed as 
\be
\kappa \eta =
 V \int^\infty_0 dp \frac{ {\kappa^\prime}^3 {p^\prime}^3 \e^{-{\kappa^\prime}^2 {p^\prime}^2}}{\sqrt{1+{p^\prime}^2} + \sqrt{1- V^2}},  
 \nonumber 
\ee
where $\kappa^\prime= m \kappa$ and $p^\prime= \p / m$. 
As shown in Appendix~\ref{sec:J_ratio}, 
$\kappa \eta$ is a monotonically decreasing function of $\kappa^\prime = m \kappa$, 
for any given velocity $V$. This then implies that $\kappa \eta$ reaches its maximum, $\sqrt{\pi} V/4$ 
at the limit of $\kappa \rightarrow 0$. 
Therefore, for $\Delta(V)$, we obtain the following inequality. 
\be
\Delta(V) \le \lim_{\kappa\to0}\Delta(V)
=\frac{1}{1-\frac{\pi}{8} V^2}. \label{eq:deltamax}
\ee

\subsection{Quantum Fisher information matrix at the relativistic limit}
We shall analyze the relativistic limit of our result in detail. 
Firstly, from Eq.\eqref{eq:deltamax}, an upper bound for the relativistic limit of the ratio $\Delta(V)$ is given by 
\be
\Delta(1) \le \frac{1}{1-\frac{\pi}{8}} \simeq 1.647. \nonumber
\ee
This shows that the ratio is always finite. 

Next, we calculate an explicit expression for the relativistic limit of 
the SLD Fisher information matrix $\Jsrel=\lim_{V\to1}\Jslmbd$. 
This is given by
\be
\Jsrel_{jk}
 = \frac{2}{\kappa^2}
\left\{1- 2 \left[ \frac{m \kappa}{2} + \frac{\sqrt{\pi}}{4} \e^{m^2 \kappa^2} (1 - 2 m^2 \kappa^2) \, \erfc(m \kappa) \right]^2\right\}
\delta_{jk}. \nonumber
\ee
It is worth noting that the $(\Jsrel)^{-1}$ is finite even at the relativistic limit of $V \rightarrow 1$ which corresponds to that of $V \rightarrow c$ in the standard unit. 
To get a further insight into the property of $\Jsrel$, 
we consider two different limits in the spread $\kappa$ of the wave function. 
We will analyze small and large $\kappa$ limit of $(\Jsrel)^{-1}$, 
as the estimation accuracy limit is quantified by the inverse of $\Jsrel$. 

When the spread is extremely broader, $\kappa \gg 1$, 
with the help of the the asymptotic expansion of the complimentary error function $\erfc(x)$ 
(see Appendix~\ref{sec:J_ratio}), 
an approximate expression of $[(\Jsrel)^{-1}]_{11}$ is written as   
\be
[(\Jsrel)^{-1}]_{11} \simeq [(\Js)^{-1}]_{11} + \frac{1}{4 m^2}. \nonumber 
\ee
The difference between $[(\Jsrel)^{-1}]_{11}$ and $[(\Js)^{-1}]_{11}$ is only a constant given by the particle mass. 

When the spread is extremely narrower, $ \kappa \ll 1$, on the other hand, 
by using the Taylor expansion (Appendix~\ref{sec:J_ratio}), we have 
\be
[(\Jsrel)^{-1}]_{11} \simeq \frac{[(\Js)^{-1}]_{11}}{1- \frac{\pi}{8}} \simeq 1.647 \, [(\Js)^{-1}]_{11}. \nonumber 
\ee
As also seen by Eq.~\eqref{eq:deltamax}, the relativistic effect for the SLD Fisher information matrix is more prominent 
when the spread is narrower. 
%
%=====================================================================
\section{Discussion} \label{sec:Discussion}
%=====================================================================
%
\subsection{No information left in spin} \label{sec:model_spin}
We show that if the moving observer does not measure the continuous degree of freedom, the observer cannot estimate the parameter shift in the position by the following reasoning. 
In other words, there is no information left in the spin of the particle. 
Putting it differently, the Wigner rotation does not transfer the information about the parameter to the spin degree of freedom. 

Suppose that the moving observer only measures the spin of the particle. 
We take the partial trace over the momentum $\vec{p}$ 
to obtain the reduced state $\rho^\Lambda_\mathrm{spin}(\theta)= \tr_\mathrm{p} \ket{{\Psi^\Lambda(\theta)}} \bra{ {\Psi^\Lambda(\theta)}}$ for this case. 
The parametric model is then given by  
\be \label{def:model_spin}
\mathcal{M}_{\rm spin}= \{\, \rho^\Lambda_\mathrm{spin}(\theta) \, | \, \theta=(\theta_1, \theta_2) \in \mathbb{R}^2 \}.
\ee
A direct calculation with using Eq.~\eqref{eq:Psilambda} gives $\rho^\Lambda_\mathrm{spin}(\theta)$ as follows. 
\begin{align}
\rho^\Lambda_{\: \theta, \, \mathrm{spin}}
&=\int d^3p \braket{p | {\Psi^\Lambda(\theta)}} \braket{ {\Psi^\Lambda(\theta)} | p} \nonumber \\
&= \int dp^1 dp^2
\sum_{\sigma=\downarrow, \uparrow} \sum_{\sigma^\prime=\downarrow, \uparrow}F_{\theta, \, \sigma}(p^1, p^2) 
F^\ast_{\theta, \, \sigma^\prime}(p^1, p^2) \nonumber \\
& \quad \times \ket{\sigma} \bra{\sigma^\prime}.  \nonumber 
 \end{align}
From Eqs.~\eqref{eq:F1} and~\eqref{eq:F2}, we see that the integrand 
$F_{\theta, \, \sigma}(p^1, p^2) F^\ast_{\theta, \, \sigma^\prime}(p^1, p^2)$,  
($\sigma, \sigma^\prime = \downarrow, \uparrow$) does not depend on the parameter $\theta$, 
since the phases cancel each other.  
Thus, in this situation, we cannot estimate the parameter of the model \eqref{def:model_spin} at all, 
since the reduced state $\rho^\Lambda_{\: \theta, \, \mathrm{spin}}$ does not depend on the parameter. 

\subsection{Effects of the Wigner rotation}
We now discuss the effect of the Wigner rotation on estimation accuracy in our model. 
As we have seen in Sec.~\ref{sec:moving_frame}, the Wigner rotation gives 
the amplitudes of both the spin up and spin down states 
When a moving observer measures the momentum only, the observer 
ends up seeing the effect of the Wigner rotation as the mixture 
of two different pure states Eq.~\eqref{eq:Psi_pi}. 
This then gives rise to the information loss, 
as the measurement of the momentum only is not complete. 
This information loss for the moving observer is, of course, expected. 
This is because the effect of the Wigner rotation followed by the partial trace 
is a completely-positive and trace-preserving map. 
Therefore, the SLD Fisher information should decrease by the monotonicity of the quantum Fisher information. 
One of the non-trivial findings of our paper is that the explicit formula 
for this information loss as a function of the velocity of the observer. 

We further elaborate on the parametric model for a moving observer. 
The wave function of the spin up state $\psi^\Lambda_{\: \uparrow}(x)$, 
which does not exist in the rest frame, appears due to the Wigner rotation. 
The peak of the probability density $| \psi^\Lambda_{\: \uparrow}(x) |^2$ no longer exists at 
$(x^{1},x^{2}=(\theta_1,\theta_2)$. 
Our numerical calculation indicates that it moves further away 
from the point $(\theta_1,\theta_2)$ as the velocity of the observer $V$ increases (Fig.~\ref{fig2} and Fig.~\ref{fig3}). 
Because of this extra peak, estimation of the expectation values of the position operators is disturbed; therefore, the SLD CR bound increases.
The ratio of the upper bound of the moving frame to the rest frame 
is given by $(1- 2 \kappa^2 \eta^2)^{-1}$, where $\eta$ is explicitly expressed as the integral form. 

Next, we comment on the role of the spread of the wave function. 
In the rest frame, $\kappa$ should be as small as possible to have better estimation accuracy. 
In Fig.~\ref{fig4}, the ratio of estimation accuracies $\Delta(V)$ is shown to be monotonically decreasing in $\kappa$ for a fixed velocity $V$. 
This means that the information loss for a moving observer is reduced by choosing relatively large $\kappa$. 
However, this results in losing estimation accuracy as the wider spread in general enables us less accurate estimation. 
Therefore, we expect the existence of a tradeoff relationship for a moving observer to design 
the best spread to gain the best information available. 
The investigation of this tradeoff will be subject to the future work. 

The relativistic limit of the SLD Fisher information is also a rather unexpected result. 
In Fig.~\ref{fig2}, we numerically evaluated the relativistic behaviors of the density for the spin up state. 
As the velocity approaches to the speed of light, we observe that the hight of the peak increases rapidly. 
This implies that the peak diverges in the relativistic limit. 
This is partially because the Lorentz transformation \eqref{eq:Lambda} does not have a well-defined limit. 
However, the SLD Fisher information matrix remains finite even in this limit, 
which is calculated by the derivatives of the state. 
Thus, the SLD CR bound does not diverge even at the relativistic limit. 

Finally, we briefly discuss achievability of the SLD CR bounds. 
We show that the SLD CR bound in the rest frame is achievable. 
When an observer is moving and does not measure the spin, the derived SLD CR bound \eqref{SLD_CRinequality_LB} is not achievable. 
This is shown by checking the weak commutativity condition~\cite{ragy, suzuki}. 
In Appendix~\ref{sec:SLD}, we calculate this condition and find that 
$ \tr(\rho_{\theta}^{\Lambda} [\Lslmbd_1 (\theta), \Lslmbd_2 (\theta)])= 8 \I \xi \eta^2 \neq 0$. 
Therefore, the SLD CR bound in the moving frame is not achievable even asymptotically. 
A further investigation of asymptotically and non-asymptotically achievable bounds shall be 
presented in due course. 
%
%=====================================================================
\section{Conclusion} \label{sec:Conclusion}
%=====================================================================
We obtain the accuracy limit for estimating the expectation value of the position 
of a relativistic particle for an observer moving along one direction at a constant velocity. 
We evaluate estimation accuracy of the position by the SLD CR bound. 
Estimation accuracy is degraded by increasing the observer's velocity. We see that this is because 
the spin up state appears in the moving frame while there exists the spin down state in the rest frame. 
Furthermore, it stays finite even at the relativistic limit. 
Since we show that the SLD CR bound is not achievable, 
it is not guaranteed that a tight CR bound gives a finite bound at the relativistic limit. 
However, since the Wigner rotation can be expressed as a rotation matrix that acts on a state vector, 
we expect that any divergent behavior will not arise from the result of applying the Wigner rotation 
to a state vector with a finite spread. 
To confirm that finiteness of estimation accuracy at the relativistic limit, 
it is important to obtain an achievable bound as a future work. 
%
%=====================================================================
\section*{Acknowledgment}
%=====================================================================
The work is partly supported by JSPS KAKENHI Grant Number JP21K11749. 
We would like to thank Dr.~P. Caban for useful comments on the manuscript. 
%
%=====================================================================
\appendix
%=====================================================================
\section{Wigner Rotation} \label{sec:Wigner}
For a massive particle with spin-1/2, we have the relation~\cite{halpern,weinberg}, 
\be
U(\Lambda) \ket{p, \sigma}
= \sqrt{\frac{(\Lambda p)^0}{p^0}} \sum_{\sigma^\prime=\downarrow, \uparrow} D^{(\frac{1}{2})}_{\sigma^\prime, \sigma} (W(\Lambda, p)) \ket{\Lambda p, \sigma},
\label{eq:Ulambdap}
\ee
where $W(\Lambda,p)=L^{-1}(\Lambda p) \Lambda L(p)$.
The Lorentz boost $L(p) =[ L^i_{\: j}(p) ]$ is chosen as in~\cite{weinberg}. 
\begin{align}
L^i_{\: j} (p) &=\delta_{i j} + \frac{(\sqrt{m^2+ \p^{\, 2}} -m) p^i p^j}{ m \p^{\, 2}}, \nonumber \\
L^i_{\: 0} (p) &=\frac{p^i}{m}, \nonumber \\
L^0_{\: 0} (p) &=\frac{\sqrt{m^2+ \p^{\, 2}}}{m}. \nonumber
\end{align}
A direct calculation for our setting $\vec{p}=(p^1, p^2, 0)$ gives the explicit representation of the matrix $W(\Lambda,p)$ as follows. 
\begin{align}
[W(\Lambda, p)]^0_{\: \: 0}&=1, \nonumber \\
[W(\Lambda, p)]^1_{\: \: 0}&=[W(\Lambda, p)]^0_{\: \: 1}=0, \nonumber \\
[W(\Lambda, p)]^2_{\: \: 0}&=[W(\Lambda, p)]^0_{\: \: 2}=0, \nonumber \\
[W(\Lambda, p)]^3_{\: \: 0}&=[W(\Lambda, p)]^0_{\: \: 3}=0, \nonumber
\end{align}
\begin{widetext}
\be
[W(\Lambda, p)]^1_{\: \: 1}=[W(\Lambda, p)]^2_{\: \: 2}
=\frac{ p^0 [m (p^1)^2 + p^0 (p^2)^2] \sinh ^2\chi + \p^2 [ (p^1)^2 \cosh \chi+(p^2)^2]}
{\p^2 \left[(p^0)^2 \sinh ^2\chi+\p^2 \right]},  \nonumber
\ee
\end{widetext}
\begin{align}
[W(\Lambda, p)]^2_{\: \: 1}&=[W(\Lambda, p)]^1_{\: \: 2} 
 = - \frac{p^1 p^2 ( \cosh \chi  -1) (p^0-m)}{\p^2 (p^0 \cosh \chi + m)},  \nonumber \\
[W(\Lambda, p)]^3_{\: \: 1}&=-[W(\Lambda, p)]^1_{\: \: 3}
=- \frac{p^1  \sinh \chi}{p^0 \cosh \chi + m}, \nonumber \\
[W(\Lambda, p)]^3_{\: \: 2}&=-[W(\Lambda, p)]^2_{\: \: 3} 
= - \frac{p^2  \sinh \chi}{p^0 \cosh \chi + m},  \nonumber \\
[W(\Lambda, p)]^3_{\: \: 3}
&=\frac{p^0 + m \cosh \chi}{m + p^0 \cosh \chi}. \nonumber
\end{align}
A $3\times3$ real matrix $[R(\Lambda, p)]_{j k}$ defined by the spatial part of $W(\Lambda, p)$ as 
\be
[R(\Lambda, p)]_{jk}= [W(\Lambda, p)]^j_{\: \: k} \, , \quad (j,k=1,2,3). \nonumber
\ee
This is a real rotation matrix acting on the three-dimensional vector space. 
We next decompose the rotation matrix $R(\Lambda, p)$ with the Euler angles. 
A straightforward calculation shows that we need only two Euler angles in this case. 
The matrices $R_2(\alpha)$ and $R_3(\phi)$ that express a rotation by angles $\alpha$ and $\phi$ around the 2 and 3-axis, 
respectively~\cite{Sakurai}, i.e.,
\be
R(\Lambda, p)
= R_3(-\phi) R_2( \alpha) R_3( \phi),  \label{eq:EulerRotation}
\ee
where
\begin{align}
R_2( \alpha)&=\begin{pmatrix}
\cos \alpha & 0 & -\sin \alpha \\
0 & 1 & 0 \\
\sin \alpha & 0 & \cos \alpha 
\end{pmatrix}, \nonumber \\
R_3( \phi)&=\begin{pmatrix}
\cos \phi & -\sin \phi & 0 \\
\sin \phi & \cos \phi & 0 \\
0 & 0 & 1 
\end{pmatrix}. \nonumber
\end{align}
As we have the Euler rotation representation, Eq.~\eqref{eq:EulerRotation}, 
we obatain the $2\times2$ matrix representation of the rotation 
for the spin-1/2 particle~\cite{Sakurai}, $D^{(\frac{1}{2})}(W(\Lambda,p))$ as
\begin{align}
D^{(\frac{1}{2})}(W(\Lambda,p))&= \e^{ \I \phi \frac{\sigma_3}{2}}  \e^{- \I \alpha \frac{\sigma_2}{2}}  \e^{ -\I \phi \frac{\sigma_3}{2}}\\
&= \begin{pmatrix}
\cos \frac{\alpha}{2}  & - \e^{\I \phi} \sin \frac{\alpha}{2} \\
 \e^{-\I \phi} \sin \frac{\alpha}{2} & \cos \frac{\alpha}{2} 
\end{pmatrix}. \nonumber
\end{align}
By substituting the expression of $D^{(\frac{1}{2})}(W(\Lambda,p))$ in ~\eqref{eq:Ulambdap}, 
we obtain Eqs.~\eqref{eq:Psilambda},~\eqref{eq:Psi_pi},~\eqref{eq:F1}, and~\eqref{eq:F2}. 
%
 %%%%%%%%%%%%
 \section{Inner product $\braket{\psi^\Lambda_{\: \sigma}(\theta) | \psi^\Lambda_{\: \sigma}(\theta)}$}  \label{sec:xi}
 %%%%%%%%%%%%
 %
 From Eq.~\eqref{eq:Psi_pi}, $\braket{\psi^\Lambda_{\: \sigma}(\theta) | \psi^\Lambda_{\: \sigma}(\theta)}$ is calculated as 
 \be
 \begin{split}
 \braket{\psi^\Lambda_{\: \sigma}(\theta) | \psi^\Lambda_{\: \sigma}(\theta)}
 &= \int d^3p \int d^3p^\prime \sqrt{\frac{(\Lambda p)^0}{p^0}} F^\ast_{\theta, \, \sigma}({p}^1, {p}^2) \delta(p^3) \\
& \quad \times  \sqrt{\frac{(\Lambda {p^\prime})^0}{{p^\prime}^0}} 
F_{\theta, \, \sigma}({{p^\prime}}^1, {{p^\prime}}^2) \delta({p^\prime}^3) \braket{\Lambda \vec{p} | \Lambda \vec{{p^\prime}}} \\
&=  \int \int | F^\ast_{\theta, \, \sigma}({p}^1, {p}^2) |^{2} dp^1 dp^2. 
\end{split} \nonumber 
 \ee
We use the relation~\cite{weinberg}, 
 \be
 \braket{\Lambda \vec{p} | \Lambda \vec{{p^\prime}}}
 = \dfrac{p^0}{(\Lambda p)^0} \braket{\vec{p} | \vec{{p^\prime}}}
 =  \dfrac{p^0}{(\Lambda p)^0} \delta(\vec{p} - \vec{p^\prime}). \nonumber
 \ee
 By using Eqs.~\eqref{eq:F1},~\eqref{eq:F2},~\eqref{eq:cosbeta}, and~\eqref{eq:sinbeta}, 
 we obtain Eqs.~\eqref{eq:rho_1} and~\eqref{eq:rho_2},~\eqref{eq:xi_min}, i.e., 
\begin{align}
\braket{\psi^\Lambda_{\: \downarrow}(\theta) | \psi^\Lambda_{\: \downarrow}(\theta)}=\frac{1}{2}( 1 + \xi), \nonumber \\
\braket{\psi^\Lambda_{\: \uparrow}(\theta) | \psi^\Lambda_{\: \uparrow}(\theta)}=\frac{1}{2}( 1 - \xi), \nonumber
\end{align}
where
\be
\xi = 2 \kappa^2 \int_0^\infty dp \, p \, \e^{-{\kappa}^2 p^2} 
\frac{\sqrt{m^2+p^2} \sqrt{1-V^2}+m}{\sqrt{m^2+p^2}+m\sqrt{1-V^2}}.  \nonumber 
\ee
From the equation above, we see that $\xi$ is a monotonically decreasing function of $\sqrt{1-V^2}$. 
Therefore, $\xi$ is a monotonically increasing function of $V$.  When $V=1$, $\xi$ takes its minimum, $\xi_\mathrm{rel}$ 
which is evaluated as follows. 
\be
\xi_\mathrm{rel}= 2 \kappa^2 \int_0^\infty \frac{p \, \e^{-{\kappa}^2 p^2}}{\sqrt{m^2+p^2}} dp
=\sqrt{\pi } m  \kappa  e^{ m^2\kappa ^2} \erfc(m \kappa ). \nonumber 
\ee
By performing the standard gaussian integration, we see $\xi=1$ when $V=0$. 
%
%%%%%%%%%%%%
\section{Probability density of a spin-1/2 particle: x-representation} \label{sec:wavefunction}
%%%%%%%%%%%%
%
We define the wave function of a particle with up spin in coordinate representation $\psi^\Lambda_{\: \uparrow}(x)$ by 
\be
\psi^\Lambda_{\: \uparrow}(x)= \braket{x | \bar{\psi}^\Lambda_{\: \uparrow}(\theta)} \big|_{\theta=0}. \nonumber 
\ee
From Eqs.~\eqref{eq:F2} and~\eqref{eq:rho_2}, the wave function $\psi^\Lambda_{\: \uparrow}(x)$ is given by
\be
\begin{split}
\psi^\Lambda_{\: \uparrow}(x)
    &=-\sqrt{\frac{2}{1-\xi}} \int d^3p \sqrt{\frac{(\Lambda p)^0}{p^0}} \varphi_0(p^1, p^2) \e^{\I \phi(p^1, \, p^2)} \\
    & \quad  \times \sin \frac{\alpha(p)}{2} \delta(p^3) \braket{x \, | \Lambda p},
\end{split} \nonumber
\ee
where $\varphi_0(p^1, p^2)=\varphi_0(p^1)\varphi_0(p^2)$. 
By a direct calculation, we have the wave function $\ \psi^\Lambda_{\: \uparrow}(x) $ as follows.
\be
\begin{split}
\psi^\Lambda_{\: \uparrow}(x)
    &=-\sqrt{\frac{2}{1-\xi}} \frac{\kappa}{(2 \pi)^2} \sqrt{\cosh \chi} \\
    & \quad \times \int dp^1 dp^2 \e^{- \kappa^2 [(p^1)^2+(p^2)^2]+ \I \phi(p^1, \, p^2)} \\
    & \quad \times \sin \frac{\alpha(p)}{2} \e^{-\I p^1 x^1 -\I p^2 x^2 -\I \sqrt{(p^1)^2+(p^2)^2 +m^2} \sinh \chi x^3}.
\end{split} \nonumber
\ee
\section{SLD and SLD Fisher information matrix}  \label{sec:SLD}
%%%%%%%%%%%%
\subsection{SLD Fisher information matrix} 
%%%%%%%%%%%%
The state we are considering $\rho^\Lambda(\theta)$ is written by 
\be
\rho^\Lambda(\theta)=
\frac{1}{2} (1+\xi) \ket{\bar{\psi}^\Lambda_{\: \downarrow}(\theta)}\bra{\bar{\psi}^\Lambda_{\: \downarrow}(\theta)} + 
\frac{1}{2} (1- \xi) \ket{\bar{\psi}^\Lambda_{\: \uparrow}(\theta)}\bra{\bar{\psi}^\Lambda_{\: \uparrow}(\theta)}. 
\nonumber
\ee
For multi-parameter models, the SLD Fisher information matrix $\Jslmbd$ for the state $\rho^\Lambda(\theta)$ 
which is non-full rank is calculated as 
\begin{align}
\Jslmbd_{jk}
=&2 (1 + \xi) [ \mathrm{Re} \braket{\partial_j \bar{\psi}^\Lambda_{\: \downarrow}(\theta) | \partial_k \bar{\psi}^\Lambda_{\: \downarrow}(\theta)} 
\nonumber \\
&-\braket{\partial_j \bar{\psi}^\Lambda_{\: \downarrow}(\theta) | \bar{\psi}^\Lambda_{\: \downarrow}(\theta)} 
\braket{ \bar{\psi}^\Lambda_{\: \downarrow}(\theta) | \partial_k \bar{\psi}^\Lambda_{\: \downarrow}(\theta)} ]  \nonumber \\
&+2 (1 - \xi) [ \mathrm{Re} \braket{\partial_j \bar{\psi}^\Lambda_{\: \uparrow}(\theta) | \partial_k \bar{\psi}^\Lambda_{\: \uparrow}(\theta)} 
\nonumber  \\
&-\braket{\partial_j \bar{\psi}^\Lambda_{\: \uparrow}(\theta) | \bar{\psi}^\Lambda_{\: \uparrow}(\theta)} 
\braket{ \bar{\psi}^\Lambda_{\: \uparrow}(\theta) | \partial_k \bar{\psi}^\Lambda_{\: \uparrow}(\theta)} ] \nonumber \\
&-
4 (1 - \xi)(1+ \xi) \nonumber \\
&\times
  \mathrm{Re} ( \braket{ \bar{\psi}^\Lambda_{\: \uparrow}(\theta) | \partial_j \bar{\psi}^\Lambda_{\: \downarrow}(\theta)}^\ast
\braket{\bar{\psi}^\Lambda_{\: \uparrow}(\theta) | \partial_k \bar{\psi}^\Lambda_{\: \downarrow}(\theta)} ). 
\label{eq: Jjk}
\end{align}
Regarding the calculation, see for example~\cite{jliu2}. 
The terms below appear in the second and fourth terms of Eq.~\eqref{eq: Jjk} vanish, because their integrands are 
an odd function of $p^j$, i.e.,
\be
\braket{\partial_j \bar{\psi}^\Lambda_{\: \sigma}(\theta) |  \bar{\psi}^\Lambda_{\: \sigma}(\theta)} = 0, 
\quad (\sigma= \downarrow, \uparrow). \nonumber
\ee

From Eqs.~\eqref{eq:Psi_pi},~\eqref{eq:F1}, and~\eqref{eq:F2}, the inner products 
$\braket{\partial_j \bar{\psi}^\Lambda_{\: \sigma}(\theta) | \partial_j \bar{\psi}^\Lambda_{\: \sigma}(\theta)}$, 
$(j=1,2)$ are obtained as follows.
\begin{align}
\braket{\partial_j \bar{\psi}^\Lambda_{\: \downarrow}(\theta) | \partial_{j} \bar{\psi}^\Lambda_{\: \downarrow}(\theta)} 
&=\frac{(2 \kappa^2)^{-1}+ \nu}{1 + \xi}, \nonumber \\
\braket{\partial_j  \bar{\psi}^\Lambda_{\: \uparrow}(\theta) | \partial_{j} \bar{\psi}^\Lambda_{\: \uparrow}(\theta)}
&=\frac{(2 \kappa^2)^{-1}- \nu}{1 - \xi}, \nonumber
\end{align}
where
\be
\nu= \int_{-\infty}^\infty \int_{-\infty}^\infty dp^1 dp^2 (p^1)^2 [\varphi_0 (p^1, p^2)]^2 \cos \alpha(\p). \nonumber
\ee
We also use Eq.~\eqref{eq:psi_0}
\be
\varphi_0(p^j)=
\frac{\kappa^{1/2}}{\pi^{1/4}}\, \e^{- \frac12\kappa^2 (p^j)^2}, \nonumber
\ee
and 
\be
\int \int dp^1 dp^2 (p^1)^2 [\varphi_0 (p^1,p^2)]^2=\frac{1}{2 \kappa^2}. \nonumber
\ee
As for $\braket{\partial_j \bar{\psi}^\Lambda_{\: \downarrow}(\theta) | \bar{\psi}^\Lambda_{\: \uparrow}(\theta)} 
\, (j=1,2)$, a direct calculation gives
\begin{align}
\braket{\partial_1 \bar{\psi}^\Lambda_{\: \downarrow}(\theta) | \bar{\psi}^\Lambda_{\: \uparrow}(\theta)}
&= - \frac{\I \eta}{ \sqrt{(1+\xi)(1-\xi)}}, \nonumber \\
\braket{\partial_2 \bar{\psi}^\Lambda_{\: \downarrow}(\theta) | \bar{\psi}^\Lambda_{\: \uparrow}(\theta)}
&= - \frac{\eta}{ \sqrt{(1+\xi)(1-\xi)}}, \nonumber
\end{align}
where 
\be
\eta=- \int_{-\infty}^\infty \int_{-\infty}^\infty dp^1 dp^2 \frac{(p^j)^2}{\p} [\varphi_0 (p^1, p^2)]^2 \sin \alpha(\p). \label{eq:eta2}
\ee

The SLD Fisher information matrix $\Jslmbd$ is expressed as follows.
\be
\Jslmbd=2 ( \kappa^{-2}-2 \eta^2)
\begin{pmatrix}
1 & 0 \\
0 & 1 \\
\end{pmatrix}. \nonumber
\ee
It turns out the $\nu$ has no effect on the SLD Fisher information. 
%
 %%%%%%%%%%%%
\subsection{SLD} 
%%%%%%%%%%%%
The SLDs, $\Lslmbd(\theta)_j$ $(j=1,2)$ are expressed by
\be
\begin{split} 
\Lslmbd_1 (\theta)
=& \frac{4}{1+ \xi}
\partial_1 ( \ket{\bar{\psi}^\Lambda_{\: \downarrow} (\theta)} ) \bra{\bar{\psi}^\Lambda_{\: \downarrow}(\theta)})  \\
&  + \frac{4}{1- \xi}
\partial_1 ( \ket{\bar{\psi}^\Lambda_{\: \uparrow} (\theta)} ) \bra{\bar{\psi}^\Lambda_{\: \uparrow}(\theta)})  \\
&  +2 \I \xi \eta
(\ket{ \bar{\psi}^\Lambda_{\: \downarrow}(\theta)}\bra{\bar{\psi}^\Lambda_{\: \uparrow}(\theta)}
- \ket{\bar{\psi}^\Lambda_{\: \uparrow}(\theta)}\bra{\bar{\psi}^\Lambda_{\: \downarrow}(\theta)} ), \nonumber
\end{split} 
\ee
\be
\begin{split}
\Lslmbd_2 (\theta)
=& \frac{4}{1+ \xi}
\partial_2 ( \ket{\bar{\psi}^\Lambda_{\: \downarrow} (\theta)} ) \bra{\bar{\psi}^\Lambda_{\: \downarrow}(\theta)}) \\
& + \frac{4}{1- \xi}
\partial_2 ( \ket{\bar{\psi}^\Lambda_{\: \uparrow} (\theta)} ) \bra{\bar{\psi}^\Lambda_{\: \uparrow}(\theta)}) \\
&+2 \xi \eta
(\ket{ \bar{\psi}^\Lambda_{\: \downarrow}(\theta)}\bra{\bar{\psi}^\Lambda_{\: \uparrow}(\theta)}
- \ket{\bar{\psi}^\Lambda_{\: \uparrow}(\theta)}\bra{\bar{\psi}^\Lambda_{\: \downarrow}(\theta)} ). \nonumber
\end{split} 
\ee
By using these,  
we can show that $\Lslmbd_1 (\theta)$ and $\Lslmbd_2 (\theta)$ do not commute, 
i.e., $[\Lslmbd_1 (\theta), \Lslmbd_2 (\theta)] \neq 0 $.

Furthermore, by a direct calculation, we can evaluate the weak commutativity condition as
\be
\tr(\rho_{\theta}^{\Lambda} [\Lslmbd_1 (\theta), \Lslmbd_2 (\theta)])= 8 \I \xi \eta^2 \neq 0. \nonumber
\ee
This shows that the SLD CR bound is not achievable even in the asymptotic setting. 
%
%%%%%%%%%%%%
\section{Maximum and minimum of $\kappa \eta$} \label{sec:J_ratio}
\begin{figure}[t]
\begin{center}
\includegraphics[width=7.5cm]{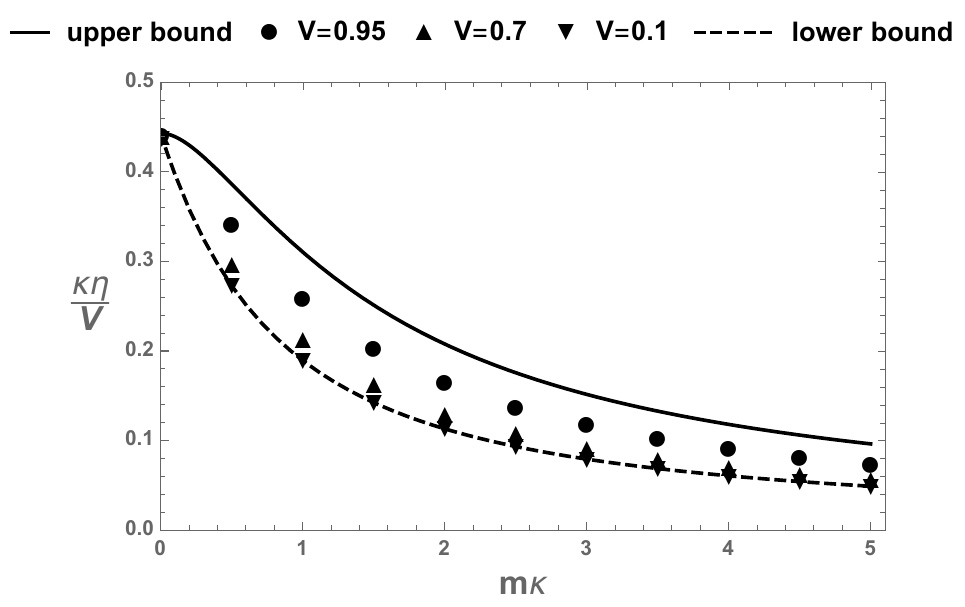}
\caption{$| \kappa \eta | / V$ as a function of $m \kappa$ at  $V=0.95, 0.7,$ and $0.1$.}
\label{fig5}
\end{center}
\end{figure}

%%%%%%%%%%%%
From Eq.~\eqref{eq:eta}, $\kappa \eta $ is expressed as 
\be
\kappa \eta =
 V \int^\infty_0 dp^\prime \frac{ {\kappa^\prime}^3 {p^\prime}^3 \e^{-{\kappa^\prime}^2 {p^\prime}^2}}{\sqrt{1+{p^\prime}^2} 
 + \sqrt{1- V^2}}. \nonumber 
\ee
By the velocity dependence of the integrand, 
we have an upper bound with $V=1$, and the lower bound with $V=0$. 
We obtain the following inequality for $\kappa \eta$. 
\be
V \int^\infty_0 dp^\prime \frac{ {\kappa^\prime}^3 {p^\prime}^3 \e^{-{\kappa^\prime}^2 {p^\prime}^2}}{\sqrt{1+{p^\prime}^2} + 1} 
\le  k \eta  \le
 V \int^\infty_0 dp^\prime \frac{ {\kappa^\prime}^3 {p^\prime}^3 \e^{-{\kappa^\prime}^2 {p^\prime}^2}}{\sqrt{1+{p^\prime}^2}}. 
 \nonumber 
\ee
These integrations are explicitly written as
\begin{align}
\int^\infty_0 dp^\prime \frac{ {\kappa^\prime}^3 {p^\prime}^3 \e^{-{\kappa^\prime}^2 {p^\prime}^2}}{\sqrt{1+{p^\prime}^2} + 1} 
&=\frac{\sqrt{\pi}}{4} \e^{{\kappa^\prime}^2} \erfc(\kappa^\prime), 
\label{eq:upper}  \\
 \int^\infty_0 dp^\prime \frac{ {\kappa^\prime}^3 {p^\prime}^3 \e^{-{\kappa^\prime}^2 {p^\prime}^2}}{\sqrt{1+{p^\prime}^2}}
&= \frac{\kappa^\prime}{2} + \frac{\sqrt{\pi}}{4} \e^{{\kappa^\prime}^2} (1 - 2 {\kappa^\prime}^2) \, \erfc(\kappa^\prime) . 
\label{eq:lower}
\end{align}
The right hand sides of Eqs.~\eqref{eq:upper}, and~\eqref{eq:lower} are monotonically decreasing functions of 
$\kappa^\prime$, or $m\kappa$.  
Their maxima at the limit of $\kappa \rightarrow 0$ for both are $\sqrt{\pi} V/4$, i.e., $\kappa \eta < \sqrt{\pi} V/4$ for any $\kappa >0$.
Figure~\ref{fig5} shows numerically calculated $| \kappa \eta | / V$ together with the upper and lower bounds. 

By using the asymptotic expansion of the complimentary error function $\erfc(x)$,
\be
\erfc(x)=\frac{\e^{-x^2}}{\sqrt{\pi} x} \sum_{n=0}^{\infty} (-1)^n \frac{(2n-1)!!}{2^n x^{2n}}, \nonumber 
\ee
for $\kappa^\prime \gg 1$, we have
\begin{align}
\int^\infty_0 dp^\prime \frac{ {\kappa^\prime}^3 {p^\prime}^3 \e^{-{\kappa^\prime}^2 {p^\prime}^2}}{\sqrt{1+{p^\prime}^2} + 1} 
& \simeq \frac{1}{4 \kappa^\prime}, \nonumber \\
 \int^\infty_0 dp^\prime \frac{ {\kappa^\prime}^3 {p^\prime}^3 \e^{-{\kappa^\prime}^2 {p^\prime}^2}}{\sqrt{1+{p^\prime}^2}}
& \simeq \frac{1}{2 \kappa^\prime}. \nonumber
\end{align} 
Then, $\Delta(V)$ is approximately expressed as
\be
1 + \frac{V^2}{8 {\kappa^\prime}^2 } \leq \Delta(V) \leq 1 + \frac{V^2}{2 {\kappa^\prime}^2}. \nonumber
\ee
For $\kappa^\prime \ll 1$, by the Taylor expansion, we have 
\be
\frac{1}{1-\frac{\pi V^2}{8}} \left( 1- \frac{\pi}{4}\frac{V^4 {\kappa^\prime}^2}{1-\frac{\pi V^2}{8}}\right) \leq \Delta(V) 
\leq \frac{1}{1-\frac{\pi V^2}{8}} \left( 1- \frac{\sqrt{\pi}}{2}\frac{V^2 {\kappa^\prime}}{1-\frac{\pi V^2}{8}}\right). \nonumber 
\ee
%
%%%%%%%%%%%%%%%


\begin{thebibliography}{15}
\bibitem{comment}
It is known that a position operator in the relativistic quantum mechanics 
is not uniquely defined. See for example Ref.~\cite{caban2} and references therein.
\end{thebibliography}
\end{document}